\documentclass[fleqn,usenatbib]{mnras}

\usepackage[T1]{fontenc}

\usepackage{graphicx}	
\usepackage{amsmath}	
\usepackage{amssymb}	
\usepackage{tabularx}
\usepackage{booktabs}
\usepackage{multirow}
\usepackage{newtxtext,newtxmath}
\usepackage{comment}
\usepackage{soul}
\usepackage[dvipsnames]{xcolor}
\usepackage{enumitem}

\def\xmm{{\it XMM-Newton}}
\def\nustar{{\it NuSTAR}}
\def\swift{{\it Swift}}

\title{A transient ultraluminous X-ray source in NGC 55}

\author[A. Robba et al.]{
A. Robba$^{1,2}$\thanks{E-mail: alessandra.robba@inaf.it},
C. Pinto$^{2}$,
F. Pintore$^{2}$,
G. Rodriguez$^{2}$,
E. Ambrosi$^{2}$,
F. Barra$^{1}$, 
G. Cusumano$^{2}$,
A. D'Aì$^{2}$,\and  
M. Del Santo$^{2}$,
P. Kosec$^{5}$, 
A. Marino$^{1,2}$,
M. Middleton$^{6}$,
T. Roberts$^{10}$,
C. Salvaggio$^{8,9}$,
R. Soria$^{3,4}$,\and
A. Wolter$^{8}$,
D. Walton$^{7,11}$
\vspace{10pt} \\
$^{1}$Universit\`a degli Studi di Palermo, Dipartimento di Fisica e Chimica, via Archirafi 36, I-90123 Palermo, Italy \\
$^{2}$INAF/IASF Palermo, via Ugo La Malfa 153, I-90146 Palermo, Italy\\
$^{3}$College of Astronomy and Space Sciences, University of the Chinese Academy of Sciences, Beijing 100049, China \\
$^{4}$Sydney Institute for Astronomy, School of Physics A28, The University of Sydney, Sydney, NSW 2006, Australia \\
$^{5}$MIT Kavli Institute for Astrophysics and Space Research, Cambridge, MA 02139, USA \\
$^{6}$Department of Physics and Astronomy, University of Southampton, Highfield, Southampton SO17 1BJ, UK \\
$^{7}$Institute of Astronomy, Madingley Road, CB3 0HA Cambridge, UK \\
$^{8}$INAF – Osservatorio Astronomico di Brera, via Brera 28, I-20121 Milano, Italy \\ 
$^{9}$Dipartimento di Fisica, Universit\`a degli Studi di Milano-Bicocca, Piazza della Scienza 3, I-20126 Milano, Italy \\
$^{10}$Centre for Extragalactic Astronomy, Department of Physics, Durham University,South Road, Durham DH1 3LE, UK \\
$^{11}$Centre for Astrophysics Research, University of Hertfordshire, College Lane, Hatfield AL10 9AB, UK
}

\date{Accepted XXX. Received YYY; in original form ZZZ}

\pubyear{2022}

\begin{document}
\label{firstpage}
\pagerange{\pageref{firstpage}--\pageref{lastpage}}
\maketitle

\begin{abstract}    
Ultraluminous X-ray sources (ULXs) are a class of accreting compact objects with X-ray luminosities above 10$^{39}$ erg s$^{-1}$. 
The average number of ULXs per galaxy is still not well constrained, especially given the uncertainty on the fraction of ULX transients. Here, we report the identification of a new transient ULX in the galaxy NGC 55 (which we label as ULX-2), thanks to recent {\xmm} and the {\it {Neil Gehrels Swift Observatory}} observations. This object was previously classified as a transient X-ray source with a luminosity around a few 10$^{38}$ erg s$^{-1}$ in a 2010 {\xmm} observation. Thanks to new and deeper observations ($\sim$ 130 ks each), we show that the source reaches a luminosity peak $>1.6 \times 10^{39}$ erg s$^{-1}$. The X-ray spectrum of ULX-2 is much softer than in previous observations and fits in the class of soft ULXs. It can be well described using a model with two thermal components, as often found in ULXs. The time scales of the X-ray variability are of the order of a month and are likely driven by small changes in the accretion rate or due to super-orbital modulations, attributed to precession of the accretion
disc, which is similar to other ULXs.
\end{abstract}




\section{Introduction}

Ultraluminous X-ray sources (ULXs) are extragalactic off-nuclear objects with bolometric X-ray luminosities $\gtrsim$ 10$^{39}$ erg s$^{-1}$, i.e. the Eddington limit for accretion onto a 10 M$_{\odot}$ black hole (e.g. \citealt{Kaaret_2017}). There is a growing consensus that ULXs are mainly powered by neutron stars (NS) or stellar-mass black holes (BHs), whose radiation is mildly beamed into our line of sight by a wind-cone in a super-Eddington regime (e.g. \citealt{Poutanen_2007}). This has been corroborated by the unambiguous discoveries of pulsations (e.g. \citealt{Bachetti_2014}) and winds (e.g. \citealt{Pinto_2016}) in a growing number of ULXs.
Other theories have suggested that some ULXs could host intermediate-mass black holes accreting at sub-Eddington rates such as for HLX-1 (\citealt{Webb2012}). 

Early studies of ULX properties in the X-ray energy band showed a particular spectral behaviour called as the \textit{ultraluminous state} \citep{Gladstone_2009}. Here, a strong spectral curvature is observed between 2--10 keV, which has been robustly confirmed by {\nustar} data (e.g. \citealt{Walton_2020}), and often coupled with a soft excess below 2 keV. Depending on the spectral slope in the 0.3--5 keV band, ULXs are generally classified as soft (SUL, $\Gamma>2$) or hard (HUL, $\Gamma<2$) ultraluminous regimes \citep{Sutton_2013}. In the latter case, if the X-ray spectrum has a single peak and a blackbody-like shape, it is classified as broadened disc regime.

The spectral properties have been interpreted in the framework of super-Eddington accretion where the disc is vertically inflated by the strong radiation pressure which is also responsible for launching winds \citep{Poutanen_2007}. Depending on the viewing angle with respect to the edge of the wind, a certain fraction of the hard photons coming from the innermost regions are obscured (edge-on) or not (face-on) by the wind \citep{Middleton_2011}. 
The different spectral shapes of ULXs and the observed switch from one regime to another (e.g. \citealt{Sutton_2013}) can be explained in terms of variability of accretion rate and wind optical depth, as well as in changing of the viewing angle \citep[e.g. due to precession,][]{Middleton_2015a}.
Such scenario can also explain the unpredictable temporal variability observed on short time-scales, i.e. from seconds to hours \citep{Heil_2009, Alston_2021}, and on longer time-scales of a few months \citep[e.g.][]{Fuerst_2018}.
Pulsations are mainly found in ULXs with hard spectra, which agrees with the overall picture and the need for a face-on configuration to detect them \citep[e.g.][]{Walton_2018}.

About 500 ULXs have been found in nearby ($d\lesssim100$ Mpc) galaxies with typically 1-2 ULXs per galaxy and a higher fraction in spirals or star forming galaxies \citep{Swartz_2011}. 
Many ULXs are transient objects, including the ones harbouring pulsating NS \citep{Hameury_2020}, whilst some high-inclination sources might be obscured by gas along the line of sight such as the Galactic super-Eddington source SS 433 \citep{Middleton_2021} and thus not appear as an ULX.
This means that the actual number of ULXs might be larger than currently thought.

In this work, we provide evidence for a new (the second) ULX in the galaxy NGC 55, thanks to recent {\xmm} observations.
The object XMMU J001446.81-391123.48 (hereafter ULX-2) was previously reported as a transient X-ray source with luminosity of a few $10^{38}$ erg s$^{-1}$ \citep{Jithesh_2015} based on archival observations with {\xmm} and the Neil Gehrels Swift Observatory (hereafter {\swift}).
Here we adopt a distance of 1.9 Mpc for the galaxy NGC 55, which is the average value of the recent Cepheid estimates\footnote{https://ned.ipac.caltech.edu/}.


\section{Observations and data analysis}

The galaxy NGC 55 was observed 10 times by {\xmm} from 2001-11-14. We do not use the early observations 0028740101-0201 as the new ULX candidate was not detected there \citep{Stobbart_2006}.
For our analysis, we particularly benefited from three recent deep ($\gtrsim 90$ ks) observations of NGC 55 (PI: Pinto). The observations were carried out with the EPIC-pn and EPIC-MOS detectors \citep{Struder_2001,Turner_2001}, all operated in full frame mode and with a thin filter. For spectral and temporal analysis we used only the data provided by pn and MOS 2, since the source fell in one of the failed MOS1 chips (CCD3 and CCD6) in some observations. 

Table \ref{XMM_observations} lists the details of the {\xmm} observations that we analysed, including the date and the net exposure for each instrument.

\begin{table}
\centering
\caption{{\xmm} observations of NGC 55.}
\renewcommand{\arraystretch}{0.9}
\small\setlength{\tabcolsep}{+2pt}
\begin{tabular}{cccccc}
\toprule
Obs.ID$(1)\atop$ & Date$(2)\atop$ & \multicolumn{2}{c}{$t_{\rm exp}$ (ks)$(3)\atop$} & \multicolumn{2}{c}{Count rate (10$^{-2}$ s$^{-1}$)} \\
& & (pn) & (MOS 2) & (pn) & (MOS 2) \\
\midrule
0655050101 & 2010-05-24 &  95.3   & 117.4 & 2.0 & 0.7 \\
0824570101 & 2018-11-17 &  90.6   & 136.5 & 21.8 & 6.9 \\
0852610101 & 2019-11-27 &   4.1   &   9.4 & 13.8 & 5.8 \\
0852610201 & 2019-12-27 &   4.1   &   6.5 & 18.6 & 6.0 \\
0852610301 & 2020-05-11 &   4.9   &   7.5 & 5.4 & 2.0 \\
0852610401 & 2020-05-19 &   4.0   &   6.5 & 14.4 & 5.5 \\
0864810101 & 2020-05-24 & 102.2   & 124.4 & 8.2 & 3.1 \\
0883960101 & 2021-12-12 & 91.3 & 117 & 13.5 & 6.6 \\
\bottomrule  
\end{tabular}

Notes:$(1)\atop$observation identifier; $(2)\atop$observation date (yyyy-mm-dd); \ $(3)\atop$net exposure times are reported after removing periods of high background. 
\label{XMM_observations}
\end{table}

   \begin{figure}
   \centering
   \includegraphics[width=8cm]{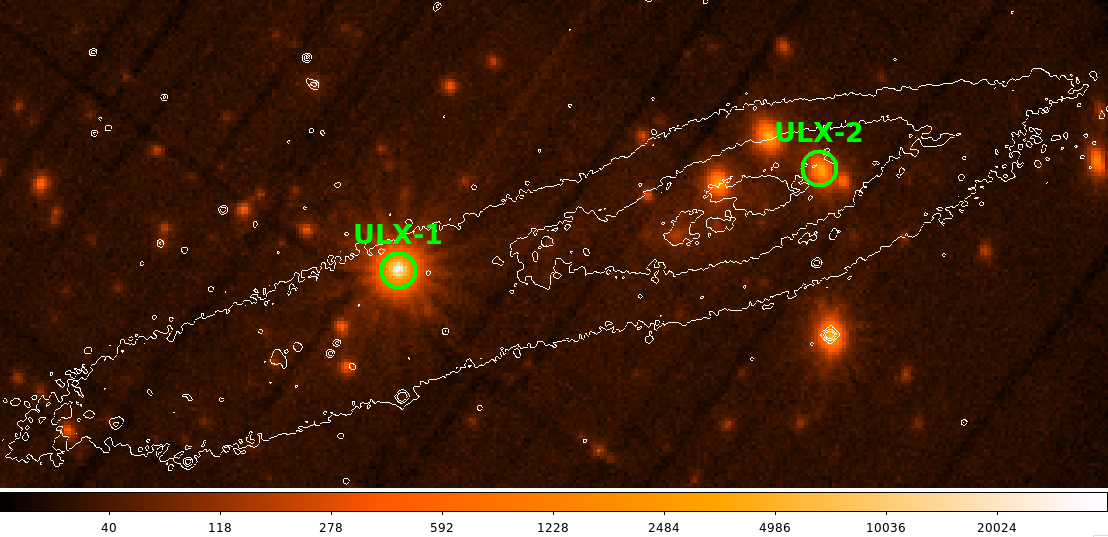}
     \vspace{-0.2cm}
      \caption{{\xmm} stacked image of NGC 55. 
      Circles with 20$^{\prime\prime}$ radii are drawn around the ULXs along with surface brightness contours from the filter-RG610 DSS image. The color bar shows the number of counts.}
    \label{image_rgb}
   \end{figure}

The data analysis was performed using the \textit{XMM-Newton Science Analysis System} (SAS) version 18.0.0 and the calibration files of January 2021.
We reduced the data from the EPIC cameras, following the standard procedure. As recommended, we selected events with FLAG=0, and PATTERN $\leq$4 ($\leq$12) for EPIC-pn (MOS), respectively. Using the task \texttt{emosaic}, we extracted the stacked EPIC MOS 1-2 and pn image in the 0.3--10\,keV  energy range. The field of the galaxy NGC 55 and the bright X-ray sources are shown in Fig.\ref{image_rgb} along with the contours derived from the archival DSS\,\footnote{https://archive.stsci.edu/cgi-bin/dss\_form} optical image.

We extracted the source and the corresponding background events from circular regions with radii of $20^{\prime\prime}$ and $50^{\prime\prime}$, respectively. The background region was chosen in a nearby region on the same chip, but away from any contaminating point source.

\subsection{Source detection}

\citet{Stobbart_2006} reported one ULX (ULX-1) in NGC 55, using the 2001 {\xmm} observations (see Fig. \ref{image_rgb}). To better classify the brightest sources located in this galaxy, we carried out a source detection on the recent deep {\xmm} data (2018, 2020 and 2021, see Table \ref{XMM_observations}).  
We ran the source detection simultaneously for EPIC-pn and EPIC-MOS for five energy bands: 0.2-0.5 keV and 0.3-0.5 keV for pn and MOS, respectively, 0.5-1 keV, 1-2 keV, 2-4.5 keV, 5.5-12 keV for both cameras. This was done with the task \texttt{edetectchain} by setting a likelihood threshold limit to the default value of 10, which corresponds to a detection threshold of $\sim$ 3 $\sigma$. 

A detailed and thorough study of the population of X-ray bright sources in NGC 55 will be reported in a forthcoming paper. Here, we report on the remarkable X-ray variability of the source XMMU J001446.81-391123.48 (ULX-2).
We detected this object in all exposures since 2010. 

\section{Spectral Analysis}

We fit the {\xmm} spectra with the {\sc{spex}} code \citep{Kaastra_1996}, over the 0.3--10 keV energy range. 
We only use the long ($\gtrsim$ 100 ks) observations, which provide sufficient statistics to decompose the spectrum. We decided to stack the spectra from ObsID 0824570101 and 0883960101 because the single spectra were compatible.
We grouped the MOS2 and pn spectra to a minimum of 25 counts per energy bin to use $\chi^2$ statistics, adopt 1\,$\sigma$ uncertainties, and fit them simultaneously. 

We tested three phenomenological models to describe the inner and outer accretion flow. In the first model we used the standard thin disc model ({\sc dbb} in {\sc spex}) and a powerlaw component ({\sc{pow}}) that are often used to model the thermal emission from the accretion disc and the Comptonised emission from the corona or the inner accretion flow, respectively (e.g. \citealt{Robba_2021}). A second model is considered to account for dominant disc emission in very bright states. This model consists of a hot disc modified by a coherent Compton scattering ({\sc mbb}, see {\sc spex} manual for more details) component and a cooler blackbody component ({\sc bb}) describing the emission from the outer disc and/or powerful winds (e.g. \citealt{Pinto_2017}).
We finally tested a combination of the standard thin disc model ({\sc dbb}) and the Comptonised emission from the corona ({\sc comt}). All continuum components are absorbed by a neutral interstellar medium, described by the {\sc hot} component (from both our Galaxy and NGC 55).

\begin{table}
\caption{Broadband continuum models.} 
\small\addtolength{\tabcolsep}{-2pt}
\resizebox{\columnwidth}{!}{
\begin{tabular}{llccc}
\toprule
Parameter & Units & 0655050101 & 0824570101+0883960101 & 0864810101 \\ 
\midrule
\multicolumn{5}{c}{Model 1 : hot (dbb + po)} \\
\midrule
$N_{\rm H}$ & (10$^{21}$/cm$^2$) & \multicolumn{3}{c}{ 6.9$+0.3\atop{-0.2}$  $(a)\atop$} \\
Norm$_{\rm dbb} $ & (10$^{16}$ m$^2$)        & 1.5$+4\atop{-1.1}$ $\times$ 10$^{-3}$ & 2.8$+1.1\atop{-0.6}$ $\times$ 10$^{-4}$ & 2.1$+15\atop{-1.8}$ $\times$ 10$^{-3}$ \\
Norm$_{\rm po}$ & (10$^{16}$ m$^2$)          & 419$+24\atop{-23}$ & 1741$+93\atop{-82}$ & 1332$+52\atop{-48}$ \\
$kT_{\rm dbb}$ & (keV)                  & 0.18$+0.03\atop{-0.02}$ & 0.37 $\pm$ 0.02 & 0.17 $\pm$ 0.04 \\
$\Gamma$ &                          & 2.90$+0.07\atop{-0.06}$ & 1.91 $\pm$ 0.04 & 2.34 $\pm$ 0.03 \\
L$_{\rm dbb,0.3-10}$ $(b)\atop$         & (10$^{38}$ erg/s)  & 1.7$+5\atop{-1.3}$ & 15$+6\atop{-3}$ & 2.2$+15\atop{-1.9}$ \\
L$_{\rm po,0.3-10}$ $(b)\atop$          & (10$^{38}$ erg/s) & 2.17 $\pm$ 0.12 & 10.5$+0.6\atop{-0.5}$ & 6.8$+0.3\atop{-0.2}$  \\	
L$_{{\rm tot},0.3-10}$         & (10$^{38}$ erg/s) & 3.9$+3\atop{-0.8}$ & 25.5$+0.5\atop{-0.2}$ & 15.0$+7\atop{-0.9}$ \\
$\chi^2$/d.o.f. &                 & 103.53/84 & 304.22/251 & 237.38/178 \\ 
\bottomrule
\multicolumn{5}{c}{Model 2 : hot (bb + mbb)} \\
\toprule
$N_{\rm H}$ & (10$^{21}$/cm$^2$) & \multicolumn{3}{c}{3.79$+0.15\atop{-0.14}$ $(a)\atop$} \\ 
Norm$_{\rm bb} $ & (10$^{16}$ m$^2$)           & 2.7$+3\atop{-1.5}$ $\times$ 10$^{-4}$ & 1.10$+0.17\atop{-0.13}$ $\times$ $^{-3}$ & 6.3$+2\atop{-1.5}$ $\times$ 10$^{-5}$ \\ 
Norm$_{\rm mbb}$ & (10$^{16}$ m$^2$)           & 0.33$+0.04\atop{-0.05}$ & 0.33$+0.03\atop{-0.02}$ & 0.45 $\pm$ 0.05 \\ 
$kT_{\rm bb}$ & (keV)                    & 0.16$+0.03\atop{-0.02}$ & 0.246 $\pm$ 0.006 & 0.31 $\pm$ 0.03 \\
$kT_{\rm mbb}$ & (keV)                   & 0.86 $\pm$ 0.04 & 2.33 $\pm$ 0.09 & 1.43 $\pm$ 0.06 \\
L$_{\rm bb,0.3-10}$ ${(b)\atop}$ & (10$^{38}$ erg/s)  & 0.13$+0.14\atop{-0.07}$ & 3.8$+0.6\atop{-0.5}$ & 0.62$+0.2\atop{-0.15}$ \\
L$_{\rm mbb,0.3-10}$ ${(b)\atop}$ & (10$^{38}$ erg/s) & 0.65$+0.10\atop{-0.08}$ & 5.8$+0.5\atop{-0.4}$ & 2.7 $\pm$ 0.3 \\
L$_{{\rm tot},0.3-10}$ & (10$^{38}$ erg/s) & 0.8$+1.2\atop{-0.7}$ & 9.6 $\pm$ 0.2 & 3.3 $\pm$ 0.4 \\
$\chi^2$/d.o.f.&            & 84.97/84 & 360.76/251 & 187.72/178 \\ 
\bottomrule
\multicolumn{5}{c}{Model 3 : hot (dbb + comt)} \\
\toprule
$N_{\rm H}$ & (10$^{21}$/cm$^2$) & \multicolumn{3}{c}{5.8 $\pm$ 0.2 $(a)\atop$} \\
Norm$_{\rm dbb}$ & (10$^{16}$ m$^2$)           & 3.7$+3\atop{-1.5}$ $\times$ 10$^{-5}$ & 5.9$+1.7\atop{-1.3}$ $\times$ 10$^{-5}$ & 1.0 $+0.5\atop{-0.3}$ $\times$ 10$^{-5}$ \\
Norm$_{\rm comt}$ & (10$^{44}$ph/(s keV))          & 118$+83\atop{-72}$ & 180$+144\atop{-148}$ & 262$+164\atop{-233}$ \\
$kT_{\rm dbb}$ & (keV)                         & 0.33 $\pm$ 0.03 &  0.49 $\pm$ 0.02  & 0.52 $+0.05\atop{-0.04}$  \\
$kt_{\rm e}$ & (keV)                           & 2.9$+19\atop{-1.1}$ & 10$+24\atop{-4}$ & 3.5$+17\atop{-1.3}$  \\
$\tau$ &                                      & \multicolumn{3}{c}{2.9$+1.3\atop{-3}$ $(a)\atop$} \\
L$_{\rm dbb,0.3-10}$ ${(b)\atop}$ & (10$^{38}$ erg/s) & 0.9$+0.7\atop{-0.4}$ & 10$+3\atop{-2}$ & 2.2$+1.1\atop{-0.6}$ \\
L$_{\rm comt,0.3-10}$ ${(b)\atop}$ & (10$^{38}$ erg/s) & 0.7$+0.5\atop{-0.4}$ & 6 $\pm$ 5 & 3$+1.7\atop{-2}$    \\
L$_{{\rm tot},0.3-10}$ & (10$^{38}$ erg/s) & 1.6$+1.5\atop{-1.1}$ & 16$+1.1\atop{-1.0}$ & 5.2$+1.1\atop{-0.9}$ \\
$\chi^2$/d.o.f.&          & 83.05/84 & 310.68/251 & 182.50/178 \\ 
\bottomrule
\bottomrule
\end{tabular}}
Notes:$(a)\atop$$N_{\rm H}$ and $\tau$ are coupled between the observations.$(b)\atop$L$_X$ (0.3--10 keV) unabsorbed luminosities are calculated assuming a distance of 1.9 Mpc.
\label{bestfit_model}
\end{table}

\begin{figure}
\includegraphics[width=9cm]{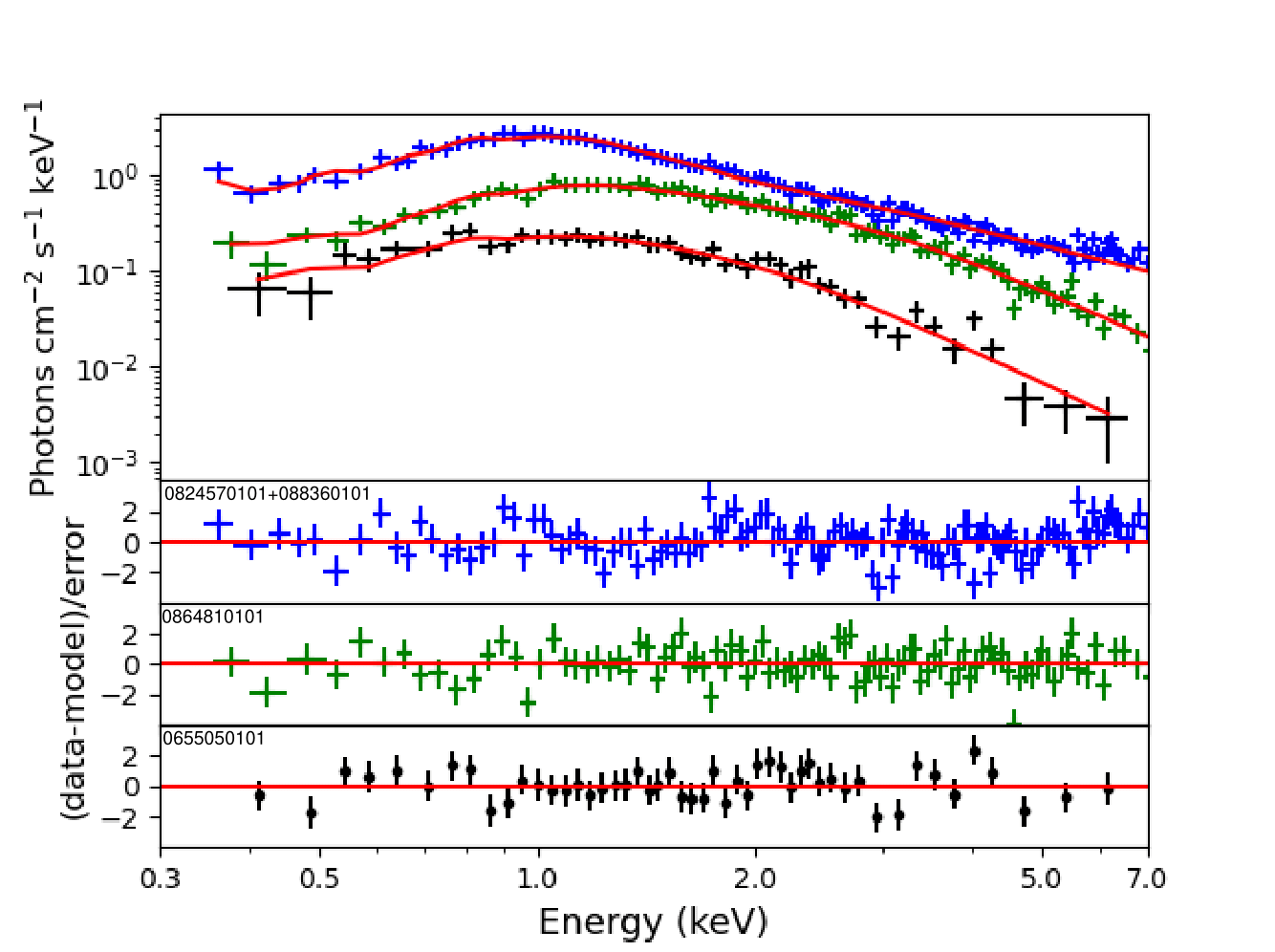}
\vspace{-0.5cm}
\caption{{\xmm} spectra of NGC 55 ULX-2 candidate. For visual purposes, only EPIC-pn data are shown. The best-fit {\sc dbb+comt} model is overlapped (red, solid line). The bottom panels show the corresponding residuals.}
\label{model_rhbmb}
\end{figure}

Since it is not yet clear whether the source is always sub-/super-Eddington or it shifts between different regimes we simultaneously applied the models to the three spectra coupling the parameters that are weakly constrained or that could vary due to some modelling artefact such as the column density, $N_{\mathrm{H}}$, and the optical depth, $\tau$. However, given that most parameters are decoupled we do not expect any effects on their best-fit value.
We also tested a model with the column density free to vary, obtaining consistent results albeit with larger uncertainties.
The resulting value is $N_{\mathrm{H}}$ $=$ (5.8 $\pm$ 0.2) $\times$ 10$^{21}$ cm$^{-2}$ for the fit with {\sc dbb+comt} model, which is much higher than 7 $\times$ 10$^{20}$ cm$^{-2}$ (Galactic value), suggesting further absorption in agreement with the source being in the central regions of the host galaxy. 
All results are shown in Figure \ref{model_rhbmb} and Table \ref{bestfit_model}. As we can see, sometimes we found almost indistinguishable $\chi^2$. There is a degeneracy between different models that can fit the spectra equally well. We have attempted to fit the spectra individually freezing the $N_{\mathrm{H}}$ to the average value and found consistent results.

The {\sc dbb+comt} model provides the best description of the broadband spectra often yielding a much lower $\chi^2$ value than the other two models. The data and the best fit model are shown in Fig. \ref{model_rhbmb}. 
All three models result in a slight under-prediction of the data above $\sim6$ keV in the brightest ULX-like spectrum, which is likely due to a hard X-ray tail as observed in many ULXs \citep{Walton_2018}. Interestingly, the brightest state (i.e. Obs.ID 0824570101+08839601) also shows residuals at 1 keV (in emission) and between 1.2-1.5 keV (in absorption) with a pattern very similar to that observed in many ULXs and resolved in a forest of emission and absorption lines produced by winds \citep{Pinto_2016,Pinto_2017}.

In order to verify these spectral features, we fitted all spectra, modifying our baseline model {\sc dbb+comt} by addying a gaussian line. We fixed the energy centroid at 1 keV and the line width to 0 (only instrumental broadening, which is $>1000$\,km/s at 1 keV). We found a Gaussian flux of (31$+9\atop{-11}$)$\times10^{44}$phot/s for the high-flux spectrum ($\Delta\chi^2$\,=\,9), which is about $10\times$ higher than its upper limit in the other observations, where the line is not significant.

As seen in Table \ref{bestfit_model}, we estimated that during the {\xmm} Obs.ID 0824570101 and 0883960101, the source reached an unabsorbed 0.3–10 keV luminosity of 1.6 $\times$ 10$^{39}$ erg s$^{-1}$ (using model {\sc dbb+comt}), which would put the source in the low-luminosity end of the typical luminosities observed in transient ULXs. 
The spectral parameters obtained for the two observations with lower flux (Obs.ID 0655050101 and 0864810101) within the uncertainties are similar to those previously obtained for the soft state of transient BHs (e.g. \citealt{DelSanto_2008}).

In order to understand the structure of the accretion disc, we also estimated the inner radius for the {\sc dbb} component in the DBB+COMT model for each observation, using the normalisation factor with the formula $r_{\rm{in}}\,=\,\sqrt{{(\rm norm}/cos\,i)}$, where $i$ is the inclination of the disc with respect to our line of sight. The inner radius results in R$_{in}$($\cos\,i$)$^{-1/2}$ $\approx$ 600, 800, and 300 km (with uncertainties of 50-100\%), for the three spectra, respectively (0655050101, 0824570101+0883960101 and 0864810101).
As we expect for transient BHs, the source becomes softer and brighter when approaching $10^{39}$\,erg/s from Obs.ID 0655050101 to 0864810101: the radius of the disc decreases resulting in an increase of the temperature likely due to a higher but still Eddington-limited $\dot{m}$. The transition to the ULX state is instead followed by an expansion of the disc photosphere as expected from super-Eddington accretion.

\section{Temporal Analysis}

\subsection{Short-term variability}

\begin{figure*}
{\includegraphics[width=8.cm]{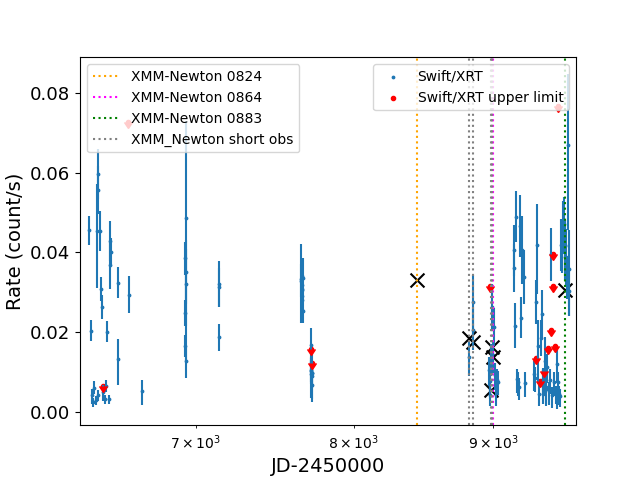}} 
{\includegraphics[width=7.1cm]{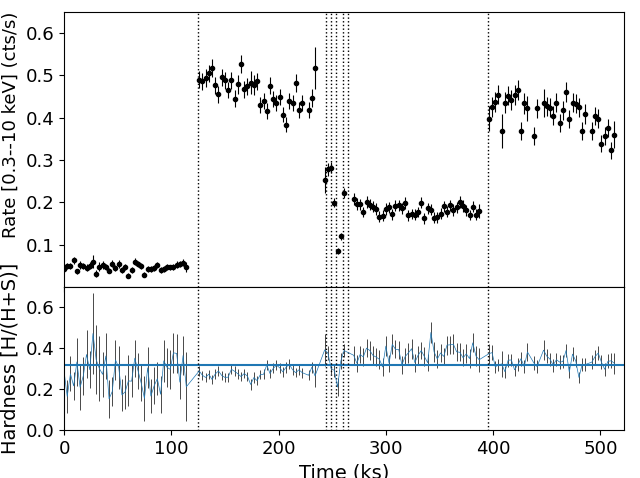}}
\caption{Left: 0.3--10 keV long-term {\swift/XRT} light curve of NGC 55 ULX-2, with the dates of the {\xmm} observations indicated by vertical dotted lines and the equivalent XRT count rates indicated by black 'X'. Right: 0.3--10 keV {\xmm} / EPIC-pn light curve of NGC 55 ULX-2. Since the observations occur at different epochs, we removed the gaps between observations with grey-dashed lines for displaying purposes.}
\label{lightcurve}
\end{figure*}

We extracted {\it XMM}/EPIC-pn light curves in the 0.3–10 keV and in two energy-selected bands (i.e. soft 0.3--1 keV and hard 1--10 keV) with a time bin of 1 ks. We computed the hardness ratios (HRs) as the ratio between the counts in the hard and the broad energy bands. We use data from all 8 observations to explore the source flux in different epochs. Due to the MOS lower count rate, we mainly use its data to confirm the overall trend measured with pn. The pn light curves of the 8 observations are then glued and shown in Fig. \ref{lightcurve} (right panel). The light curve was rebinned for visual purposes. The light curve shows $\lesssim$2\,ks dips, during which the flux decreases by a factor of 20$\%$. These dips are shallower than those seen in NGC 55 ULX-1 or other soft ULXs suggesting that a less dense wind crosses the light of sight. This is likely due to a lower accretion rate in ULX-2.

We estimate the fractional root-mean-square variability ({\it rms}) for the light curves of the three longest {\xmm} observations using standard equations \citep{Vaughan_2003}. The three light curves have comparable time baselines (115-120 ks) albeit a slightly different effective clean exposure (due to periods of high proton flaring background, see Table \ref{XMM_observations}). 
They show low intra-day variability similarly to high/soft states in XRBs (e.g. \citealt{Koljonen2018}) and many other ULXs with soft spectra ($\Gamma\sim2-3$, e.g. \citealt{Heil_2009}). The highest value ({\it rms} $\sim9$\,\%) is measured in the observation with the lowest flux. The other values are {\it rms} $\sim7$\,\% and $\sim3$\,\% for Obs.ID 0824570101 and 0864810101, respectively.

\subsection{Long-term variability}

The {\xmm} data show that the source count rate has varied by an order of magnitude over the years (and more if we account for the two earliest observations, i.e. when it was undetected). For the 2001 {\xmm} observations, \cite{Jithesh_2015} calculated the upper limit of the count rate in 0.3--8 keV, resulting in $L_{\rm X} < 10^{36}$ erg s$^{-1}$. The lack of detection might have been caused by either a variation of the disc inclination due to the precession of the accretion disc, or the onset of the propeller effect  (in the case of a magnetised neutron star), perhaps due to a lower accretion rate, below 10$^{38}$ erg s$^{-1}$, which is the lowest level found in both the {\xmm} and {\swift} light curves (see below).

The X-ray behaviour of ULX-2 is similar to that seen in GRS 1915+105, which after 25 years of high-brightness X-ray activity has decayed into a prolonged low-flux X-ray state (see \citealt{Motta_2021}). 

In order to better understand the behaviour of this source, we also studied the 0.3--10 keV historical light curve by using all the archival {\swift/XRT} observations taken between April 2013--January 2022 (see Fig. \ref{lightcurve}, left) produced with the online tool \citep{Evans_2007}\footnote{\url{https://www.swift.ac.uk/user_objects/}}. The {\swift}/XRT monitoring confirms the strong long-term variability and underlines the presence of multiple high-flux states as previously suggested by \citet{Jithesh_2015}. In Fig. \ref{lightcurve} the dotted lines show the dates of the {\xmm} observations. Unfortunately, we do not have much information on the period between 2017--2019. However, using {\scriptsize{WEBPIMMS}}\footnote{\url{https://heasarc.gsfc.nasa.gov/cgi-bin/Tools/w3pimms/w3pimms.pl}} and a simple powerlaw (with $\Gamma$=2.6) fit for the {\xmm} observation 0824570101, we predict a {\swift}/XRT count rate of 0.033 cts/s (0.3--10 keV). This corresponds to the high count-rate epochs of the XRT light curve. Moreover, we see that ULX-2 shows even higher peaks of $\sim$ 0.05 counts/sec (cts/s), which is the average {\swift/XRT} count level of NGC 55 ULX-1 \citep{Pintore_2015}.

We also investigated the possible presence of long-term periodicities, like the super-orbital variability seen in a number of ULXs, by applying a Lomb-Scargle search \citep{Lomb_1976,Scargle_1982,vanderplas_2018}. We adopted the {\sc LombScargle} function in the {\sc timeseries} class of the Python package {\sc astropy}\footnote{\url{https://docs.astropy.org/en/stable/timeseries/lombscargle.html}}.

We firstly analysed all the available {\swift}/XRT observations, choosing the background subtracted source events in the 0.3--10 keV energy band. We searched for periodicities in the frequency range (5$\times 10^{-4}$ -- 0.08) 1/d. We found that in the periodogram no statistically significant signal ($>$5$\sigma$) is found. This may be due to the scanty time coverage of the whole {\swift}/XRT dataset. 

Therefore we restricted our analyses to all the observations taken after MJD 58800 (11/2019 to 12/2021), where the {\swift}/XRT time sampling is denser. We found that the periodogram shows two peaks at periods in the range 40-60 days. The highest peak is at 41 days, but the false alarm probability (FAP) of this signal is very poor (0.81), meaning that it is not statistically significant. 

A similar approach was applied to the observations earlier than MJD 58800 (04/2013 to 12/2016).  
The highest peak was at $\sim67$ days. This could imply that during the large gap of three years the evolution of the system might have changed any super-orbital modulation. We however caution the reader that a longer and more regular monitoring of the source is necessary to confirm such a tentative periodicity.

\subsection{Search for pulsations}

We performed a deep search for periodic signals over the {\xmm} light curves, following the detection method outlined in \citet{Israel_2016}. Since significant period variations are observed in pulsating ULXs, we corrected the events time of arrival (ToA), for the presence of a first period derivative, using a grid of 3000 points, by a factor of $-\frac{1}{2}\frac{\dot{P}}{P}\,t^2$, in the range 3$\times$10$^{-6}<|\frac{\dot{P}}{P}\,($s$^{-1})|<$ 9$\times$13$^{-11}$, where $\dot{P}$ is the first period derivative. We also corrected the ToA for an orbital motion, with orbital periods in the range from 4h up to 4d and a projected semimajor axis in the range 0.05 -- 120 lt-s (see \citealt{Rodriguez_2020} for details).
No significant coherent signals were detected; we estimated a 3$\sigma$ upper limit of $\sim$ 15\% for the pulsed fraction in the data (i.e. the semi-amplitude of the signal sinusoid divided by the average count rate).

\section{Discussion and Conclusions}

In this work, we report the identification of a new ULX candidate (ULX-2) in the galaxy NGC 55. This source was previously reported as a transient X-ray source with a luminosity of a few 10$^{38}$ erg s$^{-1}$, but thanks to new, deeper, {\xmm} data, we show that the source surpasses the 10$^{39}$ erg s$^{-1}$ threshold in multiple occurrences.

It is not easy to estimate the actual number of ULXs per galaxy, due to, e.g., their variability, transient behaviour, viewing angle and local obscuration. This is important not only to understand the true form of the X-ray binary luminosity function and the binary evolution, but also to estimate the overall contribution of ULXs to galactic feedback, especially at the peak of the star formation. For all these reasons, long-term monitoring of the host galaxies is necessary to enhance the probability of detecting (specifically) transient ULXs whereas deep observations enable accurate measurements of the source spectral shape and luminosity.

We requested new observations to follow up ULX-1 and search for any new transients or strong variability in the other bright X-ray binaries in the NGC 55 galaxy. Among them, XMMU J001446.81-391123 in the past showed the most remarkable variability \citep{Jithesh_2015,Jithesh_2016}.
Considering the transient nature of some ULXs, the newly discovered ones are not always necessarily "new" sources. 
In some cases, sources are detected with luminosity lower than 10$^{39}$ erg s$^{-1}$ before reaching the ULX regime (e.g. \citealt{Hu_2018}). 

Unlike many transients, characterised by a hard spectrum with $\Gamma <2 $ (e.g. \citealt{Earnshaw_2020}), NGC ULX-2 has a soft spectrum. In the {\xmm} Obs.ID 0824570101 and 0883960101 where the source reaches $L_{X\,\rm [0.3-10\,keV]} > 10^{39}$\,erg\,s$^{-1}$, the spectrum is much softer than in previous observations and, according to \citet{Sutton_2013}, fits in the class of soft ULXs (with a $F_{\rm{pow}}/F_{\rm{dbb}}\gtrsim5$ flux ratio and $\Gamma > 2$). These sources become typically softer when brighter if the higher flux is driven by an increase in the $\dot{{M}}$ (e.g. NGC 5408 ULX-1, \citealt{Gurpide_2021a}). This is due to the fact that they are seen at moderately high inclination and at high accretion rates (close or above Eddington). The inner hot accretion flow is substantially obscured by the disc bulge around the spherisation radius and the wind. Therefore, at higher $\dot{{M}}$ the source would get even softer \citep{Middleton_2015a}. For more information on the lower states of ULX-2, see \citet{Jithesh_2015} who suggests that the source is likely a stellar-mass BH or NS XRB. A wind seems indeed to appear in the brightest (ULX) state as shown by the features observed around 1 keV (see Fig. \ref{model_rhbmb}) and formerly resolved in to emission and absorption lines from photoionised gas in many ULXs \citep[e.g.][]{Pinto_2016,Pinto_2017}. \\
The temporal behaviour shows that the time scales of the flux variations are of the order of a month. Over such period the flux can change by up to an order of magnitude. A tentative modulation on time scales of the order of 2 months appears in the earlier monitoring (2013--2016), but it is not confirmed in the follow-up observations (2019--2021) possibly due to the shorter and sparser monitoring. Super-orbital modulations on similar time scales are found in several ULXs and are sometimes attributed to precession of the accretion disc \citep{Fuerst_2018}. It is possible that, in addition to this geometric factor, the variability of the source can be associated with variations of the accretion rate. As a consequence of the increase of $\dot{m}$, the source becomes softer entering in the ULX state. Additional observations and cycles will be important to confirm or reject this result. \\
Extrapolating the best-fit model to the 0.01-100\,keV band, we estimate for the 2018 and 2021 {\xmm} observations a bolometric luminosity of $\sim 3 \times 10^{39}$\,erg\,s$^{-1}$. Its equivalent {\swift}/XRT count rate is 50\% lower than the {\swift}/XRT peaks, which suggests that the source might reach even higher bolometric luminosities, and confirms ULX-transient nature.

\section{Data availability}

All of the data and software used in this work are publicly available from ESA's {\xmm} Science Archive (XSA\footnote{https://www.cosmos.esa.int/web/xmm-newton/xsa}) and NASA's HEASARC archive \footnote{https://heasarc.gsfc.nasa.gov/}. 

\section*{Acknowledgements}
We acknowledge the anonymous referee for useful suggestions that improved the clarity of the paper. This work is based on observations obtained with {\xmm}, an ESA science mission funded by ESA Member States and USA (NASA). This work has been partially supported by the ASI-INAF programmes I/004/11/4 and 2017-14-H.0.



\bibliographystyle{mnras}
\bibliography{bibliography} 




\bsp	
\label{lastpage}

\end{document}